\documentclass{elsart}
\usepackage{amssymb,bbm}
\newcommand{\ket}[1]{\ensuremath{|#1\rangle}}
\newcommand{\bra}[1]{\ensuremath{\langle#1|}}
\newcommand{\ketbra}[1]{\ensuremath{| #1 \rangle \langle #1 |}}

\begin{document}

\begin{frontmatter}
\title{QUBIT4MATLAB V3.0: A program package for quantum information science and quantum optics for MATLAB}

\author{G\'eza T\'oth}

\address{ICFO-Institut de Ci\`encies Fot\`oniques, E-08860
Castelldefels (Barcelona), Spain}

\address{Research Institute for Solid State Physics and Optics,
Hungarian Academy of Sciences,
 P.O. Box 49, H-1525 Budapest,
Hungary}

\ead{toth@alumni.nd.edu}
\ead[url]{http://optics.szfki.kfki.hu/$\sim$toth/}

\begin{abstract}
A program package for MATLAB is introduced that helps calculations
in quantum information science and quantum optics. It has commands
for the following operations: (i) Reordering the qudits of a quantum
register, computing the reduced state of a quantum register. (ii)
Defining important quantum states easily. (iii) Formatted input and
output for quantum states and operators. (iv) Constructing operators
acting on given qudits of a quantum register and constructing spin
chain Hamiltonians. (v) Partial transposition, matrix realignment
and other operations related to the detection of quantum
entanglement. (vi) Generating random state vectors, random density
matrices and random unitaries.
\end{abstract}

% 03.65.Ud                  Entanglement and quantum nonlocality
% 03.67.Mn                  Entanglement in Quantum Information
% 03.67.-a                  Quantum information
% 03.67.Lx                  Quantum computation
% 02.70.-c                  Computational techniques
% 75.10.Pq                  Spin chain models
% 05.50.+q                  Lattice theory and statistics (Ising, Potts, etc.)

\end{frontmatter}

\newpage
\noindent{\bf Program Summary}

{\it Title of program:} QUBIT4MATLAB V3.0\\
{\it Catalogue identifier:} AEAZ\_v1\_0\\
{\it Program summary URL:}\\
http://cpc.cs.qub.ac.uk/summaries/AEAZ\_v1\_0.html,\\
also at http://arxiv.org/abs/0709.0948\\
{\it Program available from:}\\ CPC Program Library, Queen's
University, Belfast, N. Ireland, also at\\
http://optics.szfki.kfki.hu/$\sim$toth/qubit4matlab.html,\\
http://www.mathworks.com/matlabcentral/fileexchange/ \\
{\it Licensing provisions:}\\ Standard CPC licence,\\
http://cpc.cs.qub.ac.uk/licence/licence.html\\
%{\it Number of bytes in distributed program, including test code and
%documentation:} $\sim$40kB, compressed with gzip\\
{\it Programming language used:} MATLAB 6.5; runs also on Octave\\
{\it Computer:} Any which supports MATLAB 6.5.\\
{\it Operating systems:} Any which supports MATLAB 6.5; e.g.,
Microsoft Windows XP, Linux.\\
{\it Classification:} 4.15.\\

%%{\it Distribution format:} .zip\\\\

{\it Nature of Problem:}

Subroutines helping calculations in quantum
information science and quantum optics\\

{\it Method of Solution:}

A program package, that is, a set of commands is provided for
MATLAB. One can use these commands interactively or they can also be
used within a program.

\newpage

%\tableofcontents

\section{Introduction}

Quantum information science \cite{book} is one of the most rapidly
developing fields in physics. Many calculations can be done
analytically, however, many tasks need extensive numerics. In
addition, analytical calculations can very efficiently be checked
for possible errors by calculating concrete examples numerically.
The subroutine package presented in this paper was written to help
the researcher in quantum information and quantum optics in doing
such numerical calculations.

The programming effort necessary for scientific calculations in
quantum physics depends very much on the programming language used.
In particular, one has to be able to handle easily large matrices,
compute eigenvalues, eigenvectors, etc. This is certainly possible
with MATLAB, which is an interpreter language for mathematical
calculations running both under Windows and Linux. Other
alternatives may need extensive usage of complicated bracketing or
definitions of complicated data types.

The subroutine package presented is intended to fit smoothly to the
philosophy of MATLAB, and makes it possible to write down relatively
complex expressions in a concise way. Even simple functions are
defined if they are often used or their definition makes the
structure of programs clearer. After the programmer runs the main
MATLAB code, the relevant quantities, such as ground state energies
of Hamiltonians or the smallest eigenvalue of the reduced density
matrix, can be printed out writing short expressions interactively.

In this paper, the commands offered by the QUBIT4MATLAB V3.0 program
package are summarized. The first version appeared in September 2005
on the MATLAB Central File Exchange \cite{link}. Since there are
excellent books on quantum physics \cite{qmbook} and quantum
information science \cite{book}, an introduction on these topics is
not given, however, appropriate citations help the reader. The paper
is organized as follows. In Sec.~2., basic commands for defining
state vectors and density matrices are described. In Sec.~3,
commands follow that are related to reordering the qudits or tracing
out some of the qudits. In Sec.~4, definitions of interesting
quantum states, quantum gates and operators are presented. Sec.~5 is
about commands for formatted input and output. Sec.~6 lists commands
for defining two-qudit interactions and spin chain Hamiltonians.
Sec.~7 is about commands related to the separability problem. Sec.~8
is about commands using random matrices. Sec.~9 lists miscellaneous
simple commands that make programming easier. Finally, Sec.~10
summarizes commands that give sparse matrices.

The variable names most often used in the descriptions of commands
are the following:
\begin{itemize}
\item {\verb rho }: Density matrix
\item {\verb v,v1,v2,phi,phi1,phi2,psi }: State vector
\item {\verb v/rho }: A density matrix is expected. If it is not normalized,
then it is automatically normalized. If a state vector is given,
then it is converted automatically into a properly normalized
density matrix.
\item {\verb M }: Matrix
\item {\verb OP,OP1,OP2 }: Matrix corresponding to a quantum operation
\item {\verb N }: Positive integer indicating the number of qudits
\item {\verb d }: Positive integer indicating the dimension of qudits
\item {\verb k,l,m,n,k1,k2,n1,n2 }: Non-negative integer
\item {\verb list }: List of indices of qudits of a qudit register
\item {\verb perm }: List of indices indicating how to reorder the qudits (see later in detail)
\item {\verb s }: String
\end{itemize}
The square brackets {\verb [ } and {\verb ] } are used to indicate
optional parameters. If such a parameter is not given, then a
default value specific to the command is taken. In particular, for
 {\verb [N] } the default value is the value of the global variable
 {\verb N }. For
 {\verb [d] } the default value is $2$ (qubits).

\section{Bras and kets: State vectors and density matrices}

The most basic mathematical object for quantum mechanics is the
state vector. It is a vector of complex elements with unit norm. It
can be used to describe {\it pure states}. With QUIBIT4MATLAB it can
be defined with the {\verb ket } command. (Next, "bra" and "ket"
refers to the usual notation introduced by Dirac \cite{qmbook}.) For
example,
\begin{verbatim}
phi0=ket([1 0])
\end{verbatim}
defines a two element column vector as a "ket" vector with elements
$(1,0).$ In the $\{\ket{0},\ket{1}\}$ basis this corresponds to the
$\ket{\Phi_0}=\ket{0}$ state. Another example is
\begin{verbatim}
phi01=ket([1 1]).
\end{verbatim}
This defines a column vector with elements
$(\frac{1}{\sqrt{2}},\frac{1}{\sqrt{2}}),$ which corresponds to
$\ket{\Phi_{01}}=\frac{1}{\sqrt{2}}\ket{0}+\frac{1}{\sqrt{2}}\ket{1}.$
Note that {\verb ket } normalized the vector given in its argument.

The other fundamental object of quantum mechanics is the density
matrix. It is a Hermitian positive semi-definite matrix with unit
trace. Beside pure states, it can also be used to describe {\it
mixed states.} A density matrix corresponding to the previous state
vector can be defined as
\begin{verbatim}
rho=ketbra(phi01)
\end{verbatim}
If we type now {\verb rho } we obtain
\begin{verbatim}
rho =
    0.5000    0.5000
    0.5000    0.5000
\end{verbatim}
{\verb ketbra } normalizes the vector in its argument, in case it is
not normalized.

There are also further elements of the Dirac notation implemented in
QUBIT4MATLAB. One can define "bra" vectors, that is the conjugate
transpose of "ket" vectors. Hence
\begin{verbatim}
phi01b=bra([1 1])
\end{verbatim}
is a row vector with elements
$(\frac{1}{\sqrt{2}},\frac{1}{\sqrt{2}}),$ which corresponds to
$\bra{\Phi_{01}}=\frac{1}{\sqrt{2}}\bra{0}+\frac{1}{\sqrt{2}}\bra{1}.$
There is one more additional property of {\verb bra }. It computes
the complex conjugate of its argument. Thus,
\begin{verbatim}
phi01c=bra([1 i])
\end{verbatim}
will result in a row vector with elements
$(\frac{1}{\sqrt{2}},-\frac{i}{\sqrt{2}}).$

Moreover, one can define a "braket" with the command
 {\verb braket }.
\begin{verbatim}
braket(phi1,phi2)
\end{verbatim}
denotes the scalar product of two state
vectors. It is identical to {\verb bra(phi1)*ket(phi2) }. The
expression
\begin{verbatim}
braket(phi1,OP,phi2)
\end{verbatim}
where {\verb OP } is a matrix, denotes
{\verb bra(phi1)*OP*ket(phi2) }.

Finally, {\verb nm(v/rho) } normalizes its argument. If its argument
is a vector {\verb v }, then it gives back {\verb v/sqrt(v'*v) }.
This results in a unit vector. If the argument is a row vector then
it also converts it into a column vector. If the argument is a
density matrix {\verb rho } then {\verb nm } gives back
 {\verb rho/trace(rho) }. Latter results in a matrix with a unit trace.

The summary of commands for implementing the braket notation in
MATLAB and related commands are given in the following list.
\begin{itemize}
\item {\verb bra(v) }: Dirac's "bra" vector. Normalizes {\verb v } and
converts it into a column vector in case it was a row vector
\item {\verb ket(v) }: Dirac's "ket" vector. Normalizes {\verb v },
carries out an element-wise complex conjugation, and converts
 {\verb v } into a row vector in case it was a column vector
\item {\verb ketbra(v) }: Obtaining a density matrix from the state
\item {\verb ketbra2(v/rho) }: Like
 {\verb ketbra(v) }, however, a
density matrix can also be given as an argument. In this case,
{\verb ketbra2 } normalizes {\verb rho }.
\item {\verb braket(v1,v2) }: Equivalent to {\verb bra(v1)*ket(v2) }
\item {\verb braket(v1,OP,v2) }: Equivalent to {\verb bra(v1)*OP*ket(v2) }
\item {\verb ex(OP,v/rho) }: Expectation value of an operator for a
state vector or a density matrix. For normalized {\verb v } and
 {\verb rho }, it is equivalent to
 {\verb bra(v)*OP*ket(v) } and {\verb trace(OP*rho) }, respectively.
\item {\verb va(OP,v/rho) }: Variance of an operator for a
state vector or a density matrix. For normalized {\verb v } and
 {\verb rho }, it is equivalent to\\
 {\verb bra(v)*OP^2*ket(v)-(bra(v)*OP*ket(v))^2 } and \\
 {\verb trace(OP^2*rho)-trace(OP*rho)^2 }, respectively.
\item {\verb nm(v/rho) }: Normalization of a state vector or a
density matrix.
\end{itemize}

\section{Basic operations on the quantum register: Reordering qudits}

The basic object QUBIT4MATLAB handles is an array of $N$ qudits of
dimension $d.$ These qudits are numbered from $1$ to $N.$ For
example, if {\verb phi1 } and {\verb phi2 } are single-qudit state
vectors, then a two-qudit state vector can be defined as
\begin{verbatim}
phi=kron(phi2,phi1).
\end{verbatim}
This defines a product vector. The state of qubit $\#1$ is
 {\verb phi1 } and the state of qubit $\#2$ is {\verb phi2 }.
 To make it easier to handle multi-qudit registers, a Kronecker
product command with more than two arguments is defined:
 {\verb mkron }. For example,
 {\verb mkron(M1,M2,M3)=kron(kron(M1,M2),M3) }.
Moreover, there is also a "Kronecker power" function that multiplies
a matrix with itself given times using the Kronecker product. For
example, \\{\verb pkron(M,4)=kron(kron(kron(M,M),M),M) }.

Many of the following commands have {\verb N } and {\verb d } as
parameters. Typically, if {\verb d } is omitted then it is
considered to be $2,$ while if the parameter {\verb N } is omitted
then the value of the global variable {\verb N } is taken instead.
Other speciality of the commands is that at most of the places where
a density matrix is expected, a state vector can also be given. It
is automatically converted into a normalized density matrix.

Next, let us see an example. Let us define the state
$\ket{\phi}=(\ket{00}+\ket{11})\ket{1}/\sqrt{2}$ as
\begin{verbatim}
phi=ket([0 1 0 0 0 0 0 1])
\end{verbatim}
Then, we can flip the last two qudits with the command
\begin{verbatim}
phi2=reorder(phi,[3 1 2])
\end{verbatim}
When we print out {\verb phi2 }, the result is
$(\ket{010}+\ket{111})/\sqrt{2}.$ Thus the right and the middle
qubits are exchanged. In general, the second argument of
 {\verb reorder } is a list describing, how to reorder (permute) the qubits.
For $N$ qubits, $N,N-1,N-2, ... ,2,1$ corresponds to the original
configuration. Thus, the following command does not change the state
 {\verb phi }
\begin{verbatim}
phi3=reorder(phi,[3 2 1])
\end{verbatim}
The command
\begin{verbatim}
phi4=reorder(phi,[1 3 2])
\end{verbatim}
shifts the qudits cyclically to the left. When we print out
 {\verb phi4 }, the result is $(\ket{100}+\ket{111})/\sqrt{2}.$ The
meaning of the parameter describing the permutation is even
clearer if we write the numbering of qudits and the row vector
describing the permutations below each other
\begin{verbatim}
[3 2 1]
[1 3 2]
\end{verbatim}
This means that qudit $\#3$ will move to qudit $\#1,$ qudit $\#2$
will move to qudit $\#3,$ and qudit $\#1$ will move to qudit $\#2.$
The command {\verb reorder } also works for qudits with a dimension
larger than two, if a third argument is given with the dimension.

Another fundamental operation is computing the reduced density
matrix, after tracing out some of the qubits. The following
operation shows how to compute the reduced state, after tracing out
qubits $2$ and $3$
\begin{verbatim}
rho_red=remove(phi,[3 2])
\end{verbatim}
The second argument contains the list of qubits that have to be
traced out. This command also works for qudits with a dimension
larger than two, if a third argument is given. A related command is
{\verb keep }. It is essentially the same as {\verb remove }, except
that the list of the qubits that should be kept must be given.

The summary of commands for ordering/reordering qudits are given in
the following list.
\begin{itemize}
\item {\verb mkron(M1,M2,M3,...) }: Kronecker product with several arguments
\item {\verb pkron(M,n) }: Kronecker product of {\verb M } with itself {\verb n } times
\item {\verb reorder(v/rho,perm,[d]) }: Reorder the qudits of the density matrix {\verb rho } according the permutation given in {\verb perm }.
If a state vector is given instead of {\verb rho },
then the result is also a state vector.
\item {\verb reordermat(perm,[d]) }: The matrix corresponding to the quantum operation realizing a given permutation of qudits. It gives the matrix that realizes the
permutation given by {\verb perm } on a state vector of the qudit register.
\item {\verb reordervec(perm,[d]) }: The vector corresponding to the permutation of qudits. The $ith$ element of the vector tells us where to move
     the $ith$ element of a state vector during the multi-qudit register reordering.
\item {\verb shiftquditsleft(v/rho,[d]) }: Shifts the qudits of {\verb rho } to the left
\item {\verb shiftquditsright(v/rho,[d]) }: Shifts the qudits of {\verb rho } to the right
\item {\verb swapqudits(v/rho,k,l,[d]) }: Swaps the qudits {\verb k } and {\verb l } of  a quantum state {\verb rho }
\item {\verb remove(v/rho,list,[d]) }: Reduced density matrix obtained from {\verb rho }, after the qudits given in {\verb list }
are traced out
\item {\verb keep(v/rho,list,[d]) }: Reduced density matrix obtained from {\verb rho }, after the qudits {\it not} given in {\verb list }
are traced out. Thus only the qudits given in {\verb list } are
kept.
\item {\verb keep_nonorm(M,list,[d]) }:
The same as {\verb keep } but the matrix is not normalized.
\end{itemize}

\section{Definitions of important quantum states, quantum gates and operators}

There are several commands defining important quantum states and
useful operators. E.g., the simple command {\verb paulixyz } defines
the Pauli spin matrices {\verb x }, {\verb y } and
 {\verb z }. Moreover, {\verb e } is defined as the $2\times2$ identity matrix.
{\verb paulixyz } is often used in programs dealing with spin
chains. The list of such commands are:
\begin{itemize}
\item {\verb ghzstate([N]) }: State vector for the {\verb N }-qubit Greenberger-Horne-Zeilinger state
\cite{ghz}
\item {\verb wstate([N]) }: State vector for the {\verb N }-qubit
W state defined as \cite{wstate,wstate2}
\begin{equation}
\ket{W_N}=\frac{1}{\sqrt{N}}\bigg(\ket{1000...0}+\ket{0100...0}+\ket{0010...0}+...+\ket{0...0001}
\bigg).
\end{equation}

\item {\verb cstate([N]) }: State vector for the {\verb N }-qubit cluster state
\cite{cluster}. In particular, this state is the ground state of the
Hamiltonian
\begin{equation}
H_{\rm cl}(N)
:=-\sum_{k=2}^{N-1}\sigma_z^{(k-1)}\sigma_x^{(k)}\sigma_z^{(k+1)}-\sigma_x^{(1)}\sigma_z^{(2)}-\sigma_z^{(N-1)}\sigma_x^{(N)},
\end{equation}
where $\sigma_l^{(k)}$ for $l=x,y,z$ denote the Pauli spin matrices
acting on qubit $k.$
\item {\verb rstate([N]) }: State vector for the {\verb N }-qubit ring cluster state
\cite{graph}. In particular, this state is the ground state of the
Hamiltonian
\begin{equation}
H_{\rm r}(N)
:=-\sum_{k=2}^{N-1}\sigma_z^{(k-1)}\sigma_x^{(k)}\sigma_z^{(k+1)}-\sigma_z^{(N)}\sigma_x^{(1)}\sigma_z^{(2)}-\sigma_z^{(N-1)}\sigma_x^{(N)}\sigma_z^{(1)}.
\end{equation}
\item {\verb dstate(m,[N]) }: State vector for the {\verb N }-qubit symmetric Dicke state with {\verb m }
excitations \cite{dicke,josab} defined as \begin{equation}
\ket{m,N}:=\Bigg(\begin{array}{c}N \\
m\end{array}\Bigg)^{-\frac{1}{2}}\sum_k P_k
(\ket{1_1,1_2,...,1_m,0_{m+1},...,0_N}), \label{sd}
\end{equation}
where $\{P_k\}$ is the set of all distinct permutations of the
spins. $\ket{1,N}$ is the $N$-qubit W state.
\item {\verb mmstate([d],[N]) }: Density matrix of the maximally mixed state of {\verb N } qudits of dimension {\verb d }
\item {\verb mestate(d) }: State vector for the maximally entangled state of two qudits of dimension {\verb d }, that is,
\begin{equation}
\ket{\Psi_{\rm me}}:=\frac{1}{\sqrt{d}}\sum_{k=1}^d  \ket{k}\ket{k}
\end{equation}
\item {\verb singlet([N]) }: State vector for the singlet of $N$ qubits; implemented for {\verb N }$=2$ and $4.$
The two-qubit singlet is $(\ket{01}-\ket{10})/\sqrt{2}.$ The
four-qubit singlet is defined as \cite{singlet4,singlet4b}
\begin{equation}
\ket{\Phi_1}:=\frac{1}{2\sqrt{3}}\bigg(2\ket{1100}+2\ket{0011}-\ket{0101}-\ket{1010}-\ket{0110}-\ket{1001}
\bigg).
\end{equation}
\item {\verb smolinstate }: Density matrix of the state defined by Smolin
\cite{smolin}
\item {\verb gstate(Gamma) }: State vector for a graph state that was created with the Ising interaction pattern given in the
$N\times N$ adjacency matrix {\verb Gamma } \cite{graph}
\item {\verb gstate_stabilizer(Gamma) }: Gives the generators as a cell array for the stabilizer of the graph state mentioned
above \cite{graph,stab}
\item {\verb BES_Horodecki3x3(a) }: Density matrix of Horodecki's $3\times3$ bound entangled
state \cite{boundenthorodecki}. Parameter $a$ must have a value
between $0$ and $1.$
\item {\verb BES_Horodecki4x2(a) }: Density matrix of Horodecki's $4\times2$ bound entangled
state \cite{boundenthorodecki}. Parameter $a$ must have a value
between $0$ and $1.$
\item {\verb BES_UPB3x3 }: Density matrix of the $3\times3$ bound entangled state based on
unextendible product bases defined in Ref.~\cite{upb}
\item {\verb U_CNOT }: $4\times4$ unitary matrix of a CNOT gate
\item {\verb U_H }: $2\times2$ unitary matrix for the Hadamard gate
\item {\verb paulixyz }: Defines Pauli matrices {\verb x }, {\verb y },
 {\verb z } and
 {\verb e=eye(2) }
\item {\verb su3 }: Defines the $SU(3)$ generators (Gell-Mann matrices) {\verb m1 }, {\verb m2 },...,{\verb m8 } and {\verb ee } as the
$3\times3$ identity matrix \cite{su3}
\item {\verb su3_alternative }: Defines alternative $SU(3)$
generators \cite{altsu3}
\end{itemize}

\section{Formatted input and output}

The basic command for formatted output of a quantum state is
 {\verb printv }. It prints a state vector as the superposition of the
computational basis states. This function works only for qubits at
present. The form {\verb printv (v,threshold) } makes it possible to
give the threshold below which an element is considered zero. Its
usage is demonstrated on the following example
\begin{verbatim}
printv(phi2)
ans = 0.70711|010>+0.70711|111>
\end{verbatim}

The command that can be used for the formatted output of matrices is
{\verb decompose }. Its use is shown on the example
\begin{verbatim}
% Define the pauli spin matrices x,y, and z
paulixyz
% Define Heisenberg interaction for two qubits
H_H=kron(x,x)+kron(y,y)+kron(z,z)
H_H =
     1     0     0     0
     0    -1     2     0
     0     2    -1     0
     0     0     0     1
% Print out the decomposition of H_H
decompose(H_H)
ans =
xx+yy+zz
\end{verbatim}
It decomposes a Hermitian operator into the linear combinations of
products of Pauli spin matrices. Giving a second argument different
from zero makes {\verb decompose } print the results in LaTeX
format. Giving a third argument makes it possible to give the
threshold below which a coefficient is considered zero (and because
of that it is not printed).

\begin{itemize}
\item {\verb printv(v,[threshold]) }: A result is a string, giving the state vector as the superposition of multi-qubit computational basis states.
The parameter {\verb threshold } defines the limit value below which
a vector element is considered zero. If it is omitted, then it is
taken to be $10^{-4}.$
\item {\verb decompose(M,[p],[threshold]) }: The result is a string. It contains an expression describing the matrix {\verb M } as the linear combination of products of Pauli spin matrices.
If {\verb p } is not zero, then the result is given in LaTeX format.
The parameter {\verb threshold } defines the limit value below which
a coefficient is considered zero. If it is omitted then it is taken
to be $10^{-14}.$
\item {\verb paulistr(s) }: Converts a string describing an operator
constructed as a sum of products of Pauli spin matrices into an
operator. For example, {\verb op=paulistr('5*xye+xyz') } is
equivalent to {\verb paulixyz;op=5*mkron(x,y,e)+mkron(x,y,z) }. Note
that the identity is denoted by "e" for {\verb paulistr }, while it
is denoted by "1" for {\verb decompose }.
\end{itemize}

\section{Two-qudit interactions and spin chains}

When handling multi-qudit systems, one has to be able to concisely
define operators working on a given qudit. The basic command for
that is {\verb quditop(OP,k,[N]) }. It defines an  {\verb N }-qudit
quantum operator that corresponds to operator {\verb OP } acting on
the {\verb k }th qudit. Qudit position is interpreted as with
 {\verb reorder }. The dimension of the qudit is deduced from the size of
{\verb OP }. If {\verb OP } is sparse, {\verb quditop } will also
produce a sparse matrix.

Two-qudit operators can be defined by
 \verb"twoquditop(OP,k1,k2,[N])". It defines an {\verb N }-qudit quantum
operator that corresponds to the two-qudit operator {\verb OP }
acting on the
 {\verb k1 }th and {\verb k2 }th qudits. If {\verb OP } is sparse, {\verb twoquditop } will
also produce a sparse matrix. The command
 {\verb interact(OP1,OP2,n1,n2,[N]) } is an alternative way to construct an operator
acting on two qubits. It gives an operator acting on qudits
 {\verb n1 } and {\verb n2 }, respectively, with operators {\verb OP1 } and
 {\verb OP2 }. {\verb N } is the number of qudits. If argument
 {\verb N } is omitted, then the default is taken to be the value of global
variable {\verb N }. The dimension of the qudit is obtained from the
size of {\verb OP1 }.

When modeling spin chains, it is needed to construct expressions
with two-body interactions acting between nearest-neighbors. A
general form of such a nearest-neighbor interaction, for aperiodic
boundary condition, is
\begin{equation}
H_{\rm nn}(a,b,N):=\sum_{k=1}^{N-1} a^{(k)} b^{(k+1)},
\end{equation}
where $a$ and $b$ are some single-qudit operators. Their superscript
indicates on which qudit they act on. $H_{\rm nn}(a,b)$ can be
obtained by writing {\verb nnchain(a,b,N) }. The same command for
the case of periodic boundary conditions corresponding to
\begin{equation}
H_{\rm nn,p}(a,b,N):=\sum_{k=1}^{N-1} a^{(k)} b^{(k+1)}+a_Nb_1
\end{equation}
is {\verb nnchainp(a,b,N) }. When modeling spin chains, it is also
needed to define expressions of the type
\begin{equation}
H_{\rm coll}(a,N):=\sum_{k=1}^N a^{(k)},
\end{equation}
where $a$ is again a single-qudit operator. Such an expression can
be obtained writing {\verb coll(a,N) }. They are used for
defining external fields for spin chains.

After the general commands, we discuss commands specific to
particular spin chains. {\verb ising(B,[N]) } gives the
ferromagnetic Ising Hamiltonian in a transverse field
\begin{equation}
H_{\rm Ising}(B,N):=-\sum_{k=1}^{N-1} \sigma_z^{(k)}
\sigma_z^{(k+1)}+B\sum_{k=1}^N \sigma_x^{(k)}.
\end{equation}
Similarly, {\verb heisenberg(N) } gives the Heisenberg Hamiltonian defined as
\begin{equation}
H_{\rm Heisenberg}(N):=\sum_{k=1}^{N-1} \sigma_x^{(k)}
\sigma_x^{(k+1)}+\sum_{k=1}^{N-1} \sigma_y^{(k)} \sigma_y^{(k+1)}+\sum_{k=1}^{N-1}\sigma_z^{(k)}
\sigma_z^{(k+1)}.
\end{equation}
Both commands have versions for periodic boundary conditions:
 {\verb isingp(B,[N]) } and {\verb heisenbergp([N]) }. Finally, the XY
chain is in external field is defined as
\begin{equation}
H_{\rm XY}(J_x,J_y,B):=J_x\sum_{k=1}^{N-1} \sigma_x^{(k)}
\sigma_x^{(k+1)}+J_y\sum_{k=1}^{N-1}\sigma_y^{(k)} \sigma_y^{(k+1)}+B\sum_{k=1}^N
\sigma_x^{(k)}.
\end{equation}
The command giving the minimum for the XY chain for separable states
is {\verb xy_classical_ground(Jx,Jy,B) }. As the name suggests, this
minimum is the same as the ground state of the classical XY chain
\cite{xysep}.

\begin{itemize}
\item {\verb quditop(OP,k,[N]) }: Operator acting on the {\verb k }{\it th} qudit of an {\verb N }-qudit register
\item {\verb twoquditop(OP,k1,k2,[N]) }: Operator acting on qudits {\verb k1 } and {\verb k2 } of a {\verb N }-qudit register
\item {\verb coll(OP,[N]) }: Defines a collective multi-qudit operator
\item {\verb interact(OP1,OP2,n1,n2,[N]) }: Two-qudit interaction acting on qudits {\verb n1 } and {\verb n2 } of a {\verb N }-qudit register
\item {\verb nnchain(OP1,OP2,[N]) }: Spin chain Hamiltonian with a
nearest-neighbor interaction with an aperiodic boundary condition
\item {\verb nnchainp(OP1,OP2,[N]) }: Spin chain Hamiltonian with a
nearest-neighbor interaction with a periodic boundary condition
\item {\verb ising(B,[N]) }: Hamiltonian for an Ising spin chain in
a transverse field; aperiodic boundary condition
\item {\verb isingp(B,[N]) }: Hamiltonian for an Ising spin chain in
a transverse field; periodic boundary condition
\item {\verb ising_ground(B) }: Computes the ground state energy per qubit for an Ising chain in transverse field {\verb B } for the thermodynamic limit.
 The form {\verb ising_ground(B,N) } computes the same thing for an $N$-qubit chain with a periodic boundary
 condition \cite{ising}.
\item {\verb ising_free(B,T) } Free energy per qubit for an Ising chain in a transverse field {\verb B } for the thermal state for the thermodynamic limit \cite{ising}.
\item {\verb ising_thermal(B,T) }: Internal energy per spin for an Ising chain in transverse field {\verb B } for the thermodynamic limit.
The form {\verb ising_thermal(B,N) } computes the same thing for an
$N$-qubit chain with a peridodic boundary condition  \cite{ising}.
\item {\verb ising_classical_ground(B) }: Ground state energy per spin for the classical Ising
chain \cite{xysep}
\item {\verb heisenberg([N]) }: Heisenberg spin chain Hamiltonian
\item {\verb heisenbergp([N]) }: Heisenberg spin chain Hamiltonian with a periodic
boundary condition
\item {\verb xy_classical_ground(Jx,Jy,B) }: Ground state energy per spin for the classical XY
chain \cite{xysep}
\item {\verb grstate(H) }: Normalized ground state of a Hamiltonian
\item {\verb thstate(H,T) }: Thermal state of a Hamiltonian {\verb H } at temperature {\verb T }. It uses the formula $\rho_T=\exp(-H/T)/{\rm Tr}[\exp(-H/T)].$
\item {\verb orthogobs(d) }: Orthogonal observables for a qudit
with dimension {\verb d }. The orthogonal observables used are the
ones defined in Ref.~\cite{locorthog}. That is, these are the
observables of the form $\ketbra{k},$
$(\ket{k}\bra{l}+\ket{l}\bra{k})/\sqrt{2},$ or
$(\ket{k}\bra{l}-\ket{l}\bra{k})/\sqrt{2}i.$ Let us denote these
Hermitian observables by $\{M_m\}_{m=1}^{d^2}.$ They satisfy the
condition ${\rm Tr}(M_mM_n)=0$ if $m\ne n$ and ${\rm Tr}(M_m^2)=1.$
\end{itemize}

\section{Separability}

A quantum state is separable if its density matrix can be written as
the convex combination of product states, i.e., as \cite{werner}
\begin{equation}
\rho=\sum_k \rho_k^{(1)} \otimes \rho_k^{(2)} \otimes \rho_k^{(3)}
\otimes ... \otimes \rho_k^{(N)},
\end{equation}
where $N$ is the number of qudits, $p_k\ge 0$ and $\sum_k p_k=1.$ If
a quantum state is not separable, then it is entangled.

Entangled states can be used as a resource in several quantum
information processing tasks. To decide whether a state is entangled
or separable is a very important, yet, in general, unsolved question
of quantum information science. However, there are powerful
sufficient condition for entanglement in the literature, such as the
positive partial transpose (PPT) criterion \cite{ppt,ppt2} or the
computable cross norm-realignment (CCNR) criterion
\cite{ccnr,ccnr2}. For small systems we even have necessary and
sufficient conditions, and even the amount of entanglement can be
computed.

For two-qubit systems the entanglement of formation, or
equivalently, the concurrence \cite{wootters} can be computed directly from the
density matrix. This can be done with the command
 {\verb concurrence(rho) }.

A central notion is the partial transposition in quantum
information. The following command computes the partial transpose of
{\verb ketbra(phi) } with respect to the third qubit
\begin{verbatim}
rho_pt=pt(phi,3)
\end{verbatim}
In general, the second argument is a list of the indices of qudits.
The transposition will be carried out for the qubits of this list.
The command also works for qudits with dimension larger than two, if
a third argument gives the dimension. The sum of the absolute values
of the negative eigenvalues of the partial transpose is called
negativity \cite{negativity}. This can be computed by the command
 {\verb negativity }. It needs the same parameters as {\verb pt },
however, it returns a scalar value.

Beside partial transposition, there are other useful rearrangements
of the density matrix elements. Such an operation is called
realignment. For bipartite system, such a command is
 {\verb realign }. If the trace-norm of the realigned matrix is
 larger than one then the state is entangled. This can be checked by
 the {\verb ccnr } command.

There are some numerical routines looking for the maximum of an
operator for product states. This is useful for experiments: If the
operator is measured and a larger expectation value is obtained then
we know that the state is entangled. The routines are based on
simple annealing-like search for the maximum. While it is not
guaranteed that these routines find really the global maximum, they
work quite well for systems of a couple of qubits \cite{numericsep}.
For example, the maximum for separable states for a $4$-qubit
operator can be obtained
\begin{verbatim}
>> % Define the pauli spin matrices x,y, and z
paulixyz
% Define Collective operators
Jx=coll(x,4)/2; Jy=coll(y,4)/2;
% Print out the maximum for separable states
ms=maxsep(Jx^2+Jy^2)
ms =
   5.0000
\end{verbatim}
Analytical calculation shows that this is indeed the maximum for
separable states \cite{josab}. The maximum for quantum states in
general can be obtained as
\begin{verbatim}
>>  maxeig(Jx^2+Jy^2)
ans =
    6
\end{verbatim}
Thus there are quantum states for which the expectation values of
  {\verb Jx^2+Jy^2 } is larger than $5.$ These states are all
  entangled.

Finally, we briefly mention, that in a multi-qubit experiment it is
typically not enough to show that a quantum state is entangled. One
has to prove that {\it genuine multi-qubit entanglement} was present
\cite{gen_mul_ent}. It is defined as follows. If a pure state can be
written as a state separable with respect to some bipartition of the
qubits, then it is called biseparable. For example, such a state is
$(\ket{01}-\ket{10})/\sqrt{2}\otimes (\ket{01}-\ket{10})/\sqrt{2}.$
This state is the tensor product of two two-qubit singlets. While it
is entangled, it is separable with respect to the bipartition
$(12)(34).$ A mixed state is biseparable if it can be obtained by
mixing biseparable pure states. If a quantum state is not
biseparable, then it is genuine multi-qubit entangled. Several of
the commands in QUBIT4MATLAB are related to the detection of genuine
multi-qubit entanglement. For example, the maximum for biseparable
states for the previous operator can be obtained as
\begin{verbatim}
>>  maxb(Jx^2+Jy^2)
ans =
   5.2320
\end{verbatim}
Analytical calculation gives $\frac{7}{2}+\sqrt{3}\approx5.2321$
\cite{josab,dickeexp}.

The list of commands related to separability problem is summarized
in the following table.

\begin{itemize}

\item {\verb pt(v/rho,list,[d]) }: Partial transposition of {\verb rho }.
{\verb list } contains the list with indices of qudits. The qudits
on this list are transposed.

\item {\verb pt_nonorm(M,list,[d]) }: Like {\verb pt } but the matrix
given is not normalized.

\item {\verb negativity(v/rho,list,[d]) }: Negativity of {\verb rho }

\item {\verb realign(M) }: Computes the matrix obtained from realigning {\verb M }.

\item {\verb mrealign(M,iperm,[d]) }: Computes the matrix obtained from realigning the multi-qudit operator {\verb M }.
{\verb iperm } has now twice as many elements as the number of
qudits. It shows how to permute the indices of the density matrix,
if for the parties we use multiple indices. For example,
$\rho_{i_1i_2i_3,j_1j_2j_3}$ with $i_k,j_k=0,1$ would describe a
three-qubit state.

\item  {\verb cnnr(v/rho) }: Gives
directly the trace norm of the realigned matrix. A state is entangled, if the trace norm of
the realigned matrix is larger than one.

\item  {\verb optspinsq(rho) }: Optimal spin squeezing inequalities
\cite{optimspsq}. Gives back a negative value if the multi-qubit
state rho is detected as entangled by the optimal spin squeezing
inequalities. The form {\verb [fmin,f123]=optspinsq(rho) } gives
back in {\verb f123 } a three element array. Each element of the
array gives $-1$ times the violation of the corresponding spin
squeezing inequality. {\verb fmin } is the minimum of the three
values. If one of them is negative, then the state is detected as
entangled. Beside the inequalities themselves, a method is also
implemented that looks for the optimal choice of $x,$ $y,$ and $z$
coordinates. (See Ref.~\cite{optimspsq}.)

\item {\verb maxsep(OP,[d],[par]) }: Looks for
the maximum of an operator expectation value for product states. It
does not necessarily find the global maximum, but for small systems
it produces good results. {\verb d } gives the dimension of the
qudits. {\verb par } gives the parameters for the search algorithm.
It has three elements. First element: Number of random trials in the
first phase. Second element: Number of random trials in the second
phase. In the second phase the routine looks for the maximum around
the maximum found in the first phase. Third element: Constant
determining accuracy. The default value for {\verb par } is
 \verb"[ 10000 20000 0.005 ]".

\item {\verb maxsymsep(OP,[d],[par]) }: Computes the maximum only for a special case, i.e.,
for symmetric states. Because of that it is faster than
 {\verb maxsep }. {\verb d } gives the dimension of the qudits. {\verb par }
plays the same role as for {\verb maxsep }. The form
 {\verb [maximum,phi]=maxsymsep(OP) } gives back also the state giving the maximum
 in {\verb phi }. That is, the maximum is given by the state
 {\verb mkron(phi,phi,phi,...,phi) } .

\item {\verb maxbisep(OP,list,[par]) }: Gives maximum value for an operator for biseparable
states. {\verb list } determines the biparitioning. That is, the
bipartition is considered in which qubits given in {\verb list } are
in one group, the rest of the qubits are in the other group.
 {\verb par } plays the same role as for {\verb maxsep }. At the moment
works only for qubit registers.

\item {\verb maxb(OP,[par]) }: Considers the maximum for all bipartitions.
It is based on numerical optimization. {\verb par } plays the same
role as for {\verb maxsep }. It can be used, for example, when
making calculations for entanglement witnesses detecting genuine
multi-qubit entanglement in experiments. One can check with it the
bounds calculated analytically. At the moment works only for qubit
registers.

\item {\verb schmidt(v,list) }: Schmidt coefficients for a pure
state {\verb v } for the bipartition determined by {\verb list }. At
the moment works only for qubit registers.

\item {\verb overlapb(v) }: Maximum overlap with biseparable states
for a state vector {\verb v }. It is not based on numerical search,
always gives correct result. In fact, the maximum overlap is just
the square of the largest Schmidt coefficient over all bipartition
\cite{multiqubitexp}. At the moment works only for qubit registers.

\end{itemize}

\section{Commands using random matrices}

Very often it is needed to generate random state vectors, density
matrices or random unitaries. QUBIT4MATLAB has a number of commands
for these.

A random state vector (a vector of complex elements with unit
length) can be generated in the following way \cite{private}: (i)
Generate a vector such that both the real and the imaginary parts of
the vector elements are random numbers that have a normal
distribution with a zero mean and unit variance. (ii) Normalize the
vector. It is easy to prove that the random vectors obtained this
way are equally distributed on the unit sphere.

An $N$-qudit random density matrix with a distribution uniform
according to the Hilbert-Schmidt norm can be obtained in two steps
\cite{rdmat}: (i) Generate a a $2N$-qudit pure state with a
distribution uniform over the unit sphere. (ii) Trace out half of
the qudits.

Finally, an $N\times N$ random unitary with a distribution uniform
according to the Haar measure can be obtained as follows
\cite{private}: (i) Generate $N$ vectors with $N$ complex elements
and with a uniform distribution over the unity sphere. (ii)
Orthogonalize the vectors.

The list of commands using these ideas is the following:

\begin{itemize}

\item {\verb rvec([N],[d]) }: Gives a random state vector for a system of
{\verb N } qudits of dimension {\verb d }. The distribution is
uniform on the complex sphere of radius $1.$

\item {\verb rproduct([N],[d]) }: Gives the tensor product of {\verb N }
random state vectors of size {\verb d }.

\item {\verb rdmat([N],[d]) }: Gives a random density matrix for a system
of {\verb N } qudits of dimension {\verb d }. The distribution of
the matrix is uniform according to the Hilbert-Schmidt norm.

\item {\verb runitary([N],[d]) }: Gives a random unitary matrix for a
system of {\verb N } qudits of dimension {\verb d }. The
distribution of the matrix is uniform according to the Haar measure.

\item {\verb twirl(rho,[d],[Nit]) }: Twirls the multi-qudit density matrix
{\verb rho }. {\verb d } is the dimension of the qudits.
 {\verb Nit } is the number of iterations.
 The algorithm used is not simply averaging over
 random unitaries and converges very fast (for the algorithm,
 see Ref.~\cite{twirling}.)  If $d$ is omitted, then it is taken to be $2.$
  If {\verb Nit } is omitted, it is taken to be $100.$
 The form
{\verb [rho2,difference]=twirl(rho) } gives also the norm of the
difference between the original and the twirled state.  The
difference is computed through the matrix norm $\|A\|=\sum_{kl}
\|A_{kl}\|^2.$ The difference is zero for Werner states \cite{EW01}.

\item {\verb twirl2(rho,[d],[Nit]) }: Gives the maximal difference
between a multi-qudit state {\verb rho } and the state obtained from
it by a multilateral unitary rotation of the form $U\otimes U\otimes
U \otimes...\otimes U.$ The difference is computed through the
matrix norm $\|A\|=\sum_{kl} \|A_{kl}\|^2.$ {\verb d } is the
dimension of qudits. If omitted, then it is taken to be $2.$
 {\verb Nit } is the number of random unitaries used for finding the
maximum. If omitted then it is taken to be $100.$  The form
 {\verb [difference,U0]=twirl2(rho) } gives also back
    the unitary {\verb U0 } for which the difference is the largest
    between the original and the rotated state.

\end{itemize}

\section{Miscellaneous simple commands}

The following simple commands help to write programs concisely. We
discuss two of them in more detail.

{\verb trnorm(M) } gives the trace-norm of the matrix {\verb M }.
The trace-norm is defined as $\vert\vert M\vert\vert={\rm Tr
}(\sqrt{M^\dagger M}).$ It equals the sum of the singular values of
the matrix.

{\verb addnoise(v/rho,p) } adds white noise to quantum state, i.e.,
it computes
\begin{equation}
\rho'(p)=p\rho+(1-p)\frac{\mathbbm{1}}{{\rm Tr}(\mathbbm{1})}.
\nonumber
\end{equation}
The second term is the appropriately normalized identity matrix,
which corresponds to the density matrix of the completely mixed
state.

\begin{itemize}
\item \verb"proj_sym(N,[d])": Projector to the symmetric subspace
of an {\verb N }-qudit register with qudits of dimension {\verb d }.
At the moment only {\verb N }$=2$ is implemented.
\item \verb"proj_asym(N,[d])": Projector to the antisymmetric subspace
of an {\verb N }-qudit register with qudits of dimension {\verb d }.
At the moment only {\verb N }$=2$ is implemented.
\item {\verb maxeig(M) }: Maximum eigenvalue of a matrix. Defined as
{\verb max(real(eig(M))) }.
\item {\verb mineig(M) }: Minimum eigenvalue of a matrix. Defined as
{\verb min(real(eig(M))) }.
\item {\verb trace2(M) }: Trace-square of a matrix
\item {\verb trnorm(M) }: Trace-norm of a matrix
\item {\verb comm(A,B) }: Commutator, i.e., {\verb comm(A,B)=A*B-B*A }
\item {\verb addnoise(v/rho,p) }: Adds white noise to a quantum state
\item {\verb binom(m,n) }: Binomial; defined as
{\verb factorial(n)/factorial(n-m)/factorial(m) }
\item {\verb qvec([N],[d]) }: Empty state vector, filled with zeros, for {\verb N } qudits
of dimension {\verb d }
\item {\verb qsize(v/rho,[d]) }: Size of state vector or density matrix in
qudits of dimension {\verb d }
\item {\verb qeye([N],[d]) }: Identity matrix for {\verb N } qudits
of dimension {\verb d }
\end{itemize}

\section{Memory issues and commands for sparse matrices}

The amount of memory needed for storing a state vector or a density
matrix for a multi-qudit system increases exponentially with the
system size. As a rule of thumb, one can say that a $20$-qubit state
vector or a $10$-qubit density matrix is about the limit that MATLAB
can handle such that initializing these objects takes less than a
second. Going above these limits by a couple of qubits leads to "Out
of memory" error. Similarly, functions constructing spin chain
operators work also up to around $10$ qubits.

These limits can be exceeded by using sparse matrices. Sparse
matrices make it possible to store large matrices with many zero
entries very efficiently. In QUBIT4MATLAB, there are several
commands that are realized both for full matrices and for sparse
matrices. There are even commands that are only realized for sparse
matrices. They are related to two-dimensional spin systems.

Next, the sparse commands of QUBIT4MATLAB are listed. For those that
have a non-sparse version, only a short description is given.

\begin{itemize}
\item {\verb spreordermat }: Sparse version of reordermat
\item {\verb spcoll }: Sparse version of coll
\item {\verb spinteract }: Sparse version of interact
\item {\verb spnnchain }: Sparse version of nnchain
\item {\verb spnnchainp }: Sparse version of nnchainp
\item {\verb spising }: Sparse version of ising
\item {\verb spisingp }: Sparse version of isningp
\item {\verb spquditop }: Sparse version of quditop
\item {\verb sptwoquditop }: Sparse version of twoquditop
\item {\verb splatticep(op1,op2,Nx,Ny) }: Gives a two-dimensional lattice Hamiltonian
for nearest-neighbor interaction,
                          periodic boundary condition, sparse version.
                          {\verb op1 } and {\verb op2 } define the two-qudit interaction,
                          {\verb Nx } and {\verb Ny } define the
                          size of the two-dimensional lattice.
\item {\verb splattice(op1,op2,Nx,Ny) }: Gives a two-dimensional lattice Hamiltonian for nearest-neighbor interaction,
                          aperiodic boundary condition, sparse version.
                          {\verb op1 } and {\verb op2 } define the two-qudit interaction,
                          {\verb Nx } and {\verb Ny } define the
                          size of the two-dimensional lattice.
\item {\verb spising2Dp(B,Nx,Ny) }: Gives the two-dimensional ferromagnetic
Ising Hamiltonian in a transverse field, periodic boundary
condition, sparse version. {\verb B } defines the strength of the
external field. {\verb Nx } and {\verb Ny } define the size of the
two-dimensional lattice.
\end{itemize}

\section{Summary and outlook}

The QUBIT4MATLAB 3.0, a program package for MATLAB was introduced.
This package helps with the calculations in quantum information
science and quantum optics. The basic object it handles is an array
of qudits. All qudits of the array are supposed to have the same
dimension. The program package has routines for reordering the
qudits, tracing out some of the qudits, etc. It has several commands
for helping to define easily Hamilton operators for spin chains. It
has several commands related to entanglement detection, such as the
partial transposition or the realignment of the density matrix. In
future, it would be interesting to extend the routines to handle
arrays of qudits of various dimensions. This should be done without
making the notation much more complicated or making the routines
much slower.

\section{Acknowledgement}

We thank J.J. Garc\'{\i}a-Ripoll, O. G\"uhne and M.M. Wolf for
fruitful discussions. This work was supported by the Spanish MEC
(Ramon y Cajal Programme, Consolider-Ingenio 2010 project ''QOIT'')
and the EU IP SCALA. We also thank the support of the National
Research Fund of Hungary OTKA (Contract No. T049234) and the
Hungarian Academy of Sciences (J\'anos Bolyai Programme).

\end{document}